\newcommand{\be}{\begin{equation}}
\newcommand{\ee}{\end{equation}}
\newcommand{\bea}{\begin{eqnarray}}
\newcommand{\eea}{\end{eqnarray}}

\newcommand{\aeq}{&=&}

\newcommand{\itDelta}{{\it \Delta}}

\newcommand{\itLambda}{{\Lambda}}

\newcommand{\bra}{\langle}
\newcommand{\ket}{\rangle}
\newcommand{\dbra}{\bra \! \bra}
\newcommand{\dket}{\ket \! \ket}

\newcommand{\bq}{{\bar q}}

\newcommand{\rT}{{\rm T}}

\newcommand{\rR}{{\rm R}}

\newcommand{\rRe}{{\rm Re}}

\newcommand{\rS}{{\rm S}}

\documentclass{elsart}
\usepackage{epsfig}

\usepackage{amssymb}

\begin{document}

\begin{frontmatter}



\title{Multifractal analysis of fluid particle accelerations 
in turbulence}




\author[label2]{Toshihico~Arimitsu}\ead{arimitsu@cm.ph.tsukuba.ac.jp} and
\author[label3]{Naoko~Arimitsu}

\address[label2]{Institute of Physics, University of Tsukuba, Ibaraki 305-8571,
Japan}
\address[label3]{Graduate School of EIS, Yokohama Nat'l.~University, 
Yokohama 240-8501, Japan}

\begin{abstract}
The probability density function (PDF) of
accelerations in turbulence is derived analytically with the help of 
the multifractal analysis 
based on generalized entropy, i.e., the Tsallis or the R\'{e}nyi entropy.
It is shown that the derived PDF explains quite well the one obtained 
by Bodenschatz et al.\ in the measurement of fluid particle accelerations 
in the Lagrangian frame at $R_\lambda = 690$, 
and the one by Gotoh et al.\ in the DNS with the mesh size 1024$^3$ 
at $R_\lambda = 380$.
\end{abstract}

\end{frontmatter}

\section{Introduction}
\label{intro}

The {\it multifractal analysis} of turbulence \cite{AA,AA1,AA2,AA3,AA4,AA5,AA6,AA7,AA8,AA9} 
is a unified self-consistent approach for the systems with large deviations,
which has been constructed based on 
the Tsallis-type distribution function \cite{Tsallis88,Tsallis99}
that provides an extremum of the {\it extensive} R\'{e}nyi \cite{Renyi} 
or the {\it non-extensive} Tsallis entropy \cite{Tsallis88,Tsallis99,Havrda-Charvat}
under appropriate constraints.
The analysis rests
on the scale invariance of the Navier-Stokes equation for high Reynolds number, 
and on the assumptions that the singularities due to the invariance 
distribute themselves multifractally in physical space.
The multifractal analysis belongs to the line of study 
based on a kind of {\it ensemble} theoretical approaches that, 
starting from K41 \cite{K41}, continues with 
the log-normal model \cite{Oboukhov62,K62,Yaglom}, the $\beta$-model \cite{Frisch78},
the p-model \cite{Meneveau87a,Meneveau87b},
the 3D binomial Cantor set model \cite{Hosokawa91} and so on.
After a rather preliminary investigation of the p-model \cite{AA},
we developed further to derive the analytical expression for 
the scaling exponents of velocity structure function \cite{AA1,AA2,AA3,AA4}, and to
determine the probability density function (PDF) of velocity fluctuations 
\cite{AA4,AA5,AA6,AA7}, of velocity derivative \cite{AA8} and of fluid particle
accelerations \cite{AA9} by a self-consistent statistical mechanical approach.
It has been shown \cite{AA4} that
the multifractal analysis derives the log-normal model \cite{Oboukhov62,K62,Yaglom} 
when one starts with the Boltzmann-Gibbs entropy.

In this paper, we derive the formula for the PDF of the accelerations
of a fluid particle in fully developed turbulence by means of 
the multifractal analysis, and will analyze two experiments. 
One is the PDF of accelerations at $R_\lambda = 690$ of 
the Taylor microscale Reynolds number obtained in 
the Lagrangian measurement of particle accelerations 
that was realized by Bodenschatz et al.\ \cite{EB01a,EB01b,EB02comment} 
by raising dramatically the spatial and temporal measurement resolutions
with the help of the silicon strip detectors.
The other is the PDF of accelerations at $R_\lambda = 380$ \cite{Gotoh pressure} 
that was extracted by Gotoh from
the DNS of the size 1024$^3$ \cite{Gotoh02} which may be the largest
mesh size available at present.

For high Reynolds number $\rRe \gg 1$, or for the situation where 
effects of the kinematic viscosity $\nu$ can be neglected compared with
those of the turbulent viscosity, the Navier-Stokes equation,
\be
\partial {\vec u}/\partial t
+ ( {\vec u}\cdot {\vec \nabla} ) {\vec u} 
= - {\vec \nabla} \left(p/\rho \right)
+ \nu \nabla^2 {\vec u},
\label{N-S eq}
\ee
of an incompressible fluid is invariant under 
the scale transformation~\cite{Moiseev76,Frisch-Parisi83,Meneveau87b}
\be
{\vec r} \rightarrow \lambda {\vec r}, \quad
{\vec u} \rightarrow \lambda^{\alpha/3} {\vec u}, \quad
t \rightarrow \lambda^{1- \alpha/3} t, \quad 
\left(p/\rho\right) \rightarrow \lambda^{2\alpha/3} \left(p/\rho\right)
\ee
where the exponent $\alpha$ is an arbitrary real quantity.
The quantities $\rho$ and $p$ represent, respectively, mass density and pressure.
The acceleration $\vec {\mathrm{a}}$ of a fluid particle is given by
the substantive time derivative of the velocity:
\be
{\vec {\mathrm a}} = \partial {\vec u}/\partial t
+ ( {\vec u}\cdot {\vec \nabla} ) {\vec u}.
\ee
The Reynolds number $\rRe$ of the system is given by 
\be
{\rm Re} = \delta u_{\rm in} \ell_{\rm in}/\nu = ( \ell_{\rm in}/\eta )^{4/3}
\label{Re}
\ee
with the Kolmogorov scale 
$
\eta = ( \nu^3/\epsilon )^{1/4}
$~\cite{K41}
where $\epsilon$ is the energy input rate at the {\em energy-input} scale 
$\ell_{\rm in}$.
The velocity fluctuation $\delta u_{\rm in}$ is defined by
putting $\ell_n = \ell_{\rm in}$ in
$
\delta u_n = \vert u(\bullet + \ell_n) - u(\bullet) \vert
$
where $u$ is a component of velocity field $\vec{u}$, and
$\ell_n$ is a distance between two observing points.
We are measuring distance by the discrete units 
\be
\ell_n = \delta_n \ell_0
\label{r-n}
\ee
with $\delta_n = 2^{-n}$ $(n=0,1,2,\cdots)$ and $\ell_0$ being 
some reference length.
The non-negative integer $n$ represents the {\it multifractal depth}.
However, we will treat it as positive real number in the analysis of 
experiments.

It may be worthwhile to note here that, within the energy cascade model,
the diameter $\ell_{\bar{n}}$ of eddies should be defined by
\be
\ell_{\bar{n}} = \delta_{\bar{n}} \ell_{\rm in}.
\label{r-nbar}
\ee
Therefore, if one identifies the distance $\ell_n$ with the diameter $\ell_{\bar{n}}$ 
of the $\bar{n}$th eddies within the energy cascade model, one has
the relation 
\be
\bar{n} = n - \log_2 \left(\ell_0/\ell_{\rm in}\right)
\label{n-nbar}
\ee
between the number $n$ of the multifractal steps and 
the number $\bar{n}$ of the steps within the energy cascade.
We see that $\ell_{\rm in} = \ell_{\bar{0}}$ as it should be.

Introducing the {\it pressure} (divided by the mass density) difference 
\be
\delta p_n = \vert p/\rho(\bullet + \ell_n) - p/\rho(\bullet) \vert
\ee
between two points separated by the distance $\ell_n$, 
the accelerations may be estimated by 
\be
\vert \vec{\mathrm{a}} \vert = \lim_{n \rightarrow \infty} \mathrm{a}_n
\ee
where we introduced the acceleration 
\be
\mathrm{a}_n = \delta p_n / \ell_n
\ee
belonging to the multifractal depth $n$.
The acceleration becomes singular for $\alpha < 1.5$, i.e.,
\be
\lim_{n \rightarrow \infty} \mathrm{a}_n 
= \lim_{\ell_n \rightarrow 0} \delta p_n/\ell_n
\sim \lim_{\ell_n \rightarrow 0} \ell_n^{(2\alpha/3)-1}
\rightarrow \infty
\ee
which can be seen with the relation
\be
\delta p_n / \delta p_0 = (\ell_n / \ell_0)^{2\alpha/3}.
\label{p-alpha}
\ee
The values of exponent $\alpha$ specify the degree of singularity.

\section{Multifractal spectrum}

The multifractal analysis rests on the multifractal distribution of $\alpha$.
The probability 
$
P^{(n)}(\alpha) d\alpha
$
to find, at a point in physical space, a singularity labeled by an exponent 
in the range 
$
\alpha \sim \alpha + d \alpha
$
is given by~\cite{AA1,AA2,AA3,AA4}
\be
P^{(n)}(\alpha) = Z_{\alpha}^{(n)\ -1} \left[ 1 - (\alpha - \alpha_0)^2 \big/ (\itDelta \alpha )^2 
\right]^{n/(1-q)}
\label{Tsallis prob density}
\ee
with an appropriate partition function
$
Z_{\alpha}^{(n)}
$
and
\be
(\itDelta \alpha)^2 = 2X \big/ [(1-q) \ln 2 ].
\ee
The range of $\alpha$ is $\alpha_{\rm min} \leq \alpha \leq \alpha_{\rm max}$ with
\be
\alpha_{\rm min} = \alpha_0 - \itDelta \alpha, 
\quad
\alpha_{\rm max} = \alpha_0 + \itDelta \alpha.
\ee
It may be worthwhile to put here its brief derivation in order to make the paper 
self-contained. 
The distribution function (\ref{Tsallis prob density}) is 
derived by taking an extremum of the generalized entropy,
the R\'{e}nyi entropy~\cite{Renyi} 
\be
S_{q}^{\rR}[P^{(1)}(\alpha)] = \left(1-q \right)^{-1} 
\ln \int d \alpha P^{(1)}(\alpha)^{q}
\label{SqR-alpha}
\ee
or the Tsallis entropy~\cite{Tsallis88,Tsallis99,Havrda-Charvat}
\be
S_{q}^{\rT}[P^{(1)}(\alpha)] = \left(1-q \right)^{-1}
\left(\int d\alpha \ P^{(1)}(\alpha)^{q} -1 \right),
\label{SqTHC-alpha}
\ee
under the two constraints, i.e., the normalization of distribution function:
\be
\int d\alpha P^{(1)}(\alpha) = \mbox{const.},
\label{cons of prob}
\ee
and the $q$-variance being kept constant as a known quantity:
\be
\sigma_q^2 = \left(\int d\alpha P^{(1)}(\alpha)^{q} 
(\alpha- \alpha_0 )^2 \right) \Big/ \int d\alpha P^{(1)}(\alpha)^{q}.
\label{q-variance}
\ee
Here, we assume that the distribution function at the $n$th 
multifractal depth has the structure
\be
P^{(n)}(\alpha) \propto [P^{(1)}(\alpha)]^n.
\ee
This is consistent with the relation \cite{Meneveau87b,AA4}
\be
P^{(n)}(\alpha) \propto \delta_n^{1-f(\alpha)}
\ee
that is a manifestation of scale invariance and 
reveals how densely each singularity, labeled by $\alpha$, 
fills physical space.
In the present model, the multifractal spectrum $f(\alpha)$
is given by~\cite{AA1,AA2,AA3,AA4}
\be
f(\alpha) = 1 + (1-q)^{-1} \log_2 [ 1 - (\alpha - \alpha_0 )^2
/ (\Delta \alpha )^2 ].
\label{Tsallis f-alpha}
\ee
In spite of the different characteristics of these entropies,
i.e., extensive and non-extensive,
the distribution functions giving their extremum
have the common structure (\ref{Tsallis prob density}).\footnote{
Within the present formulation, the decision cannot be pronounced 
which of the entropies is underlying the system of turbulence.
}

The dependence of the parameters $\alpha_0$, $X$ and $q$ on 
the intermittency exponent $\mu$ is determined, 
self-consistently, with the help of the three independent equations, i.e.,
the energy conservation:
\be
\left\bra \epsilon_n \right\ket = \epsilon,
\label{cons of energy}
\ee
the definition of the intermittency exponent $\mu$:
\be
\bra \epsilon_n^2 \ket 
= \epsilon^2 \delta_n^{-\mu},
\label{def of mu}
\ee
and the scaling relation\footnote{
The scaling relation is a generalization of the one derived first in
\cite{Costa,Lyra98} to the case where the multifractal spectrum
has negative values.
}:
\be
1/(1-q) = 1/\alpha_- - 1/\alpha_+
\label{scaling relation}
\ee
with $\alpha_\pm$ satisfying $f(\alpha_\pm) =0$. 
The average $\bra \cdots \ket$ is taken with $P^{(n)}(\alpha)$.
The energy-transfer rate $\epsilon_n$ represents
the rate of transfer of energy per unit mass from eddies of size 
$\ell_n = \ell_{\bar{n}}$ (the $\bar{n}$th step within the energy cascade model) 
to those of size 
$\ell_{n+1} = \ell_{\overline{n+1}}$ (the $\overline{n+1}$th step within 
the energy cascade model).
It should be noted that the average $\bra \epsilon_n \ket$ is taken just for 
the eddies having the size $\ell_n$. It is {\em not} the average within 
the spatial region of the diameter $\ell_n$.

For the region $0.13 \leq \mu \leq 0.40$
where the value of $\mu$ is usually observed,
the three self-consistent equations are solved to give 
the approximate equations~\cite{AA7}:
\be
\alpha_0 = 0.9989 + 0.5814 \mu,
\quad
X = - 2.848 \times 10^{-3} + 1.198 \mu
\ee
\be
q = -1.507 + 20.58 \mu - 97.11 \mu^2 + 260.4 \mu^3 - 365.4 \mu^4 + 208.3 \mu^5.
\ee

\section{Scaling exponent of velocity structure function}

Let us derive first the probability 
$\itLambda^{(n)}(y_n) dy_n$ to find the scaled pressure fluctuations
\be
\vert y_n \vert = \delta p_n /\delta p_0
\ee
in the range $y_n \sim y_n+dy_n$ in the form
\be
\itLambda^{(n)}(y_n) dy_n = \itLambda^{(n)}_{\rS}(y_n) dy_n 
+ \Delta \itLambda^{(n)}(y_n) dy_n
\label{def of Lambda}
\ee
with the normalization
\be
\int_{-\infty}^{\infty} dy_n  \itLambda^{(n)}(y_n) =1.
\ee
Here, we assumed that it has two contributions whose origins are independent
with each other.
The first term represents the contribution by the singular part of accelerations 
stemmed from the multifractal distribution of the singularities 
in physical space. This is given by
\be
\itLambda^{(n)}_{\rS}(\vert y_n \vert) dy_n \propto P^{(n)}(\alpha) d \alpha
\label{singular portion}
\ee
with the transformation of the variables,
$
\vert y_n \vert = \delta_n^{2\alpha/3}
$.
Whereas the second term $\Delta \itLambda^{(n)}(y_n) dy_n$ represents 
the contribution from the dissipative term
in the Navier-Stokes equation, and/or the one 
from the errors in measurements.
The dissipative term has been discarded in the above investigation since it violates 
the invariance under the scale transformation.
The contribution of the second term provides a correction to the first one.
Note that the proportionality coefficient in (\ref{singular portion}) 
determines the portion of the contribution among these two independent 
origins.\footnote{
Needless to say that each term in (\ref{def of Lambda}) is a multiple of two PDFs, 
i.e., the PDF for one of the two independent origins to realize and the conditional PDF
for a value $y_n$ in the range $y_n \sim y_n + d y_n$ to come out.
This is of course in a generalized sense in which the second correction term
may weaken the first singular contribution.
}

The $m$th moments of the pressure fluctuations,
$
\dbra \vert y_n \vert^m \dket 
$,
are given by 
\be
\dbra \vert y_n \vert^m \dket \equiv \int_{-\infty}^{\infty} dy_n  
\vert y_n \vert^m \itLambda^{(n)}(y_n)
= 2 \tilde{\gamma}^{(n)}_m
+ (1-2\tilde{\gamma}^{(n)}_0 ) \
a_{2m} \ \delta_n^{\zeta_{2m}}
\label{structure func m}
\ee
where
\be
2\tilde{\gamma}^{(n)}_m = \int_{-\infty}^{\infty} dy_n\ 
\vert y_n \vert^m \Delta \itLambda^{(n)}(y_n)
\ee
\be
a_{3\bq} = \{ 2 / [\sqrt{C_{\bq}} ( 1+ \sqrt{C_{\bq}} ) ] \}^{1/2}
\ee
with
\be
{C}_{\bq} = 1 + 2 \bq^2 (1-q) X \ln 2.
\label{cal D}
\ee
The quantity 
\be
\zeta_m = \alpha_0 m/3 
- 2Xm^2/[9 (1+C_{m/3}^{1/2} )]
- [1-\log_2 (1+C_{m/3}^{1/2} ) ] /(1-q)
\label{zeta}
\ee
is the scaling exponent of the {m}th order velocity structure function
which explains successfully the experimental results \cite{AA1,AA2,AA3,AA4}.
Note that the formula is independent of $n$.

\section{PDF of the fluid particle accelerations}

We now derive the PDF, $\hat{\itLambda}^{(n)}(\omega_n)$, 
defined by the relation 
\be
\hat{\itLambda}^{(n)}(\omega_n) d\omega_n
= \itLambda^{(n)}(y_n) d y_n
\ee
with the {\it acceleration} $\omega_n$ normalized by its deviation, i.e.,
\be
\vert \omega_n \vert = \mathrm{a}_n / \dbra \mathrm{a}_n^2 \dket^{1/2}
= \delta p_n/\dbra \delta p_n^2 \dket^{1/2}
= y_n/\dbra y_n^2 \dket^{1/2}
= \bar{\omega}_n \delta_n^{2\alpha /3 -\zeta_4 /2}
\ee
with
\be
\bar{\omega}_n = [2 \tilde{\gamma}_2^{(n)} \delta_n^{-\zeta_4} + (1-2\tilde{\gamma}_0^{(n)} ) 
a_4 ]^{-1/2}.
\ee
It is reasonable to imagine that the origin of intermittent rare events is 
attributed to the singular term in (\ref{def of Lambda}). We then have, for
the {\em tail part} $\omega_n^\dagger \leq \vert \omega_n \vert \leq \omega_n^{\rm max}$,
\bea
\hat{\itLambda}^{(n)}(\omega_n) d \omega_n 
\aeq \itLambda^{(n)}_{\rm S} (y_n) dy_n
\nonumber\\
\aeq 
 \tilde{\itLambda}^{(n)}_{\rS} \frac{\bar{\omega}_n}{\vert \omega_n \vert}
\left[1 - \frac{1-q}{n}\ 
\frac{\left(3 \ln \vert \omega_n / \omega_{n,0} \vert\right)^2}{
8X \vert \ln \delta_n \vert} \right]^{n/(1-q)} d \omega_n
\label{PDF accel large}
\eea
with
\be
\omega_{n,0} = \bar{\omega}_n \delta_n^{2\alpha_0 /3 -\zeta_4 /2},
\quad
\omega_n^{\rm max} = \bar{\omega}_n \delta_n^{2\alpha_{\rm min}/3 -\zeta_4 /2}
\ee
and
\be
\tilde{\itLambda}^{(n)}_{\rS} = 3 \left(1-2\tilde{\gamma}^{(n)}_0 \right)
\Big/ \left(4 \bar{\omega}_n \sqrt{2\pi X \vert \ln \delta_n \vert} \right).
\ee
On the other hand, for smaller accelerations, 
the contribution to the PDF comes from both the singularity and thermal fluctuations 
or measurement error. We assume that this part of the PDF is described by 
Tsallis-type function with a new parameter $q'$, i.e.,
for the {\em center part} 
$
\vert \omega_n \vert \leq \omega_n^\dagger
$,
\bea
\lefteqn{
\hat{\itLambda}^{(n)}(\omega_n) d \omega_n =
\left[ \hat{\itLambda}^{(n)}_{\rS}(y_n)
+\Delta \hat{\itLambda}^{(n)}(y_n) \right] d y_n }
\nonumber\\
\aeq 
\tilde{\itLambda}^{(n)}_{\rS}
\left\{1-\frac{1-q'}{2} \left[1+\frac{3}{2}f'(\alpha^\dagger) \right]
\left[ \left(\frac{\omega_n}{\omega_n^\dagger}\right)^2 -1 \right] \right\}^{1/(1-q')}
 d \omega_n.
\label{PDF accel small}
\eea
This specific form of the Tsallis function is determined by the condition that 
the two PDFs (\ref{PDF accel large}) and (\ref{PDF accel small}) 
should have the same value and the same slope at
$\omega_n^\dagger$ which is defined by
\be
\omega_n^\dagger = \bar{\omega}_n \delta_n^{2\alpha^\dagger /3 -\zeta_4 /2}
\ee
with $\alpha^\dagger$ being the smaller solution of 
\be
\zeta_4/2 -2\alpha/3 +1 -f(\alpha) = 0.
\ee
It is the point at which 
$
\hat{\itLambda}^{(n)}(\omega_n^\dagger)
$
has the least $n$-dependence for large $n$.

With the help of the relation (\ref{PDF accel small}),
we obtain $\Delta \itLambda^{(n)}(y_n)$, and have the formula 
to evaluate $\tilde{\gamma}_m^{(n)}$ in the form
\be
2\tilde{\gamma}_m^{(n)} = \left(K_m^{(n)} - L_m^{(n)}\right) \Big/
\left(1 + K_0^{(n)} - L_0^{(n)}\right)
\ee
where
\bea
K_m^{(n)} \aeq \frac{3\ \delta_n^{2(m+1)\alpha^\dagger/3 -\zeta_4/2}}
{2\sqrt{2 \pi X \vert \ln \delta_n \vert}}
\int_0^1 dz\ z^{m} 
\nonumber\\
&& \hspace*{2cm} \times
\left[1-\frac{1-q'}{2} \left[1+\frac{3}{2}f'(\alpha^\dagger)\right]
\left( z^2 -1 \right) \right]^{1/(1-q')},
\\
L_m^{(n)} \aeq \frac{3\ \delta_n^{2m \alpha^\dagger/3}}
{2\sqrt{2 \pi X \vert \ln \delta_n \vert}}
\int_{z_{\rm min}}^1 dz\ z^{m-1} 
\left[1 - \frac{1-q}{n}\ \frac{\left(3 \ln \vert z / z_0^\dagger \vert \right)^2}{
8X \vert \ln \delta_n \vert} \right]^{n/(1-q)}
\eea
with
\be
z_{\rm min} = \omega_{\rm min}/\omega_n^\dagger 
=\delta_n^{2(\alpha_{\rm max} - \alpha^\dagger)/3},
\quad
z_0^\dagger = \omega_{n,0}/\omega_n^\dagger 
=\delta_n^{2(\alpha_0 - \alpha^\dagger)/3}.
\ee

Now, the PDF of fluid particle accelerations, given by 
(\ref{PDF accel large}) and (\ref{PDF accel small}), is completely determined by 
three parameters, i.e., 
the intermittency exponent $\mu$, the multifractal depth $n$
which gives a characteristic length $\ell_n$, and 
$q'$ which appears in the Tsallis-type PDF at the center part.
The intermittency exponent $\mu$ is determined by
analyzing the measured scaling exponent $\zeta_m$ of the velocity structure function
with the formula (\ref{zeta}) as mentioned before.

\begin{figure}
\begin{center}
\includegraphics[width=7cm]{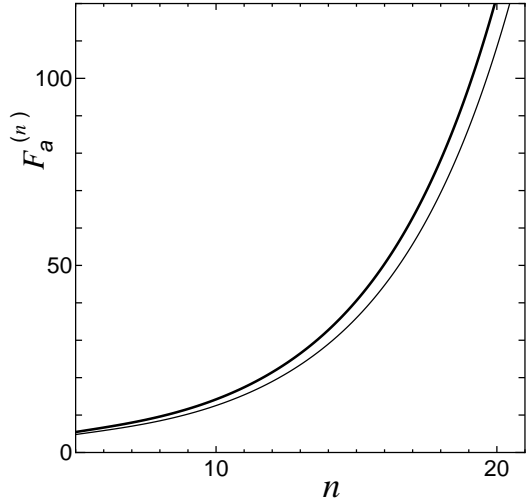}
\end{center}
\caption{Dependence of the flatness on the multifractal step $n$.
The thin and thick lines represent, respectively, the dependence 
for the parameters used in the study of the experiment by Bodenschatz et al.\
(Fig.~\ref{PDF acceleration log-linear Bodenschatz} below),
and of the DNS by Gotoh et al.\ (Fig.~\ref{PDF acceleration log-linear Gotoh} below).
\label{fig: F-n}}
\end{figure}

As for the determination of the value $n$, the flatness of the PDF of fluid particle 
accelerations can be a good candidate \cite{EB02comment}.
Actually, the present multifractal analysis provides us with the analytical formula
for the flatness of the PDF in the form
\bea
F_{\mathrm{a}}^{(n)} &\equiv& \dbra \mathrm{a}_n^4 \dket
/\dbra \mathrm{a}_n^2 \dket^2
=\dbra \omega_n^4 \dket 
= \frac{2 \tilde{\gamma}^{(n)}_4
+ (1-2\tilde{\gamma}^{(n)}_0 ) \
a_{8} \ \delta_n^{\zeta_{8}}}
{\left[2 \tilde{\gamma}^{(n)}_2
+ (1-2\tilde{\gamma}^{(n)}_0 ) \
a_{4} \ \delta_n^{\zeta_{4}}\right]^2}
\label{F-n}
\\
&\approx& \frac{a_{8}}
{(1-2\tilde{\gamma}^{(n)}_0 ) \
a_{4}^2} \ \delta_n^{\zeta_{8}-2\zeta_{4}}.
\label{F-n approx}
\eea
With this formula, the experimental value of the flatness $F_{\mathrm{a}}^{(n)}$ 
gives the value of the multifractal step $n$. 
In deriving the approximate formula (\ref{F-n approx}), we used the fact that 
the first terms both in the denominator and numerator of the last formula in (\ref{F-n})
are two or three orders in magnitude smaller than the second terms.
Note that the contribution to the flatness is mainly come from
the tail part of the PDF, and that there is almost no contribution from 
its center part (see Fig.~\ref{a4Lambda} below).
Therefore, at this stage, we can determine almost the final shape
and the magnitude of the tail part given by (\ref{PDF accel large}).
It means that the shape of the center part, which is controlled by $q'$, 
rarely affect the tail part, and that the contribution of the center part
to the normalization of PDF is almost independent of $q'$.
The $n$-dependence of the flatness is given in Fig.~\ref{fig: F-n} 
both for the cases corresponding to the experiments conducted by Bodenschatz et al.\ 
with the approximate formula
\bea
F_{\mathrm{a}}^{(n)} \aeq -6.872 + 5.216 n - 9.969 \times 10^{-1} n^2
\nonumber\\
&&+ 1.075 \times 10^{-1} n^3 - 5.296 \times 10^{-3} n^4 
+ 1.241 \times 10^{-4} n^5,
\eea
and to the DNS performed by Gotoh et al.\ with 
\bea
F_{\mathrm{a}}^{(n)} \aeq -7.592 + 5.831 n - 1.114 n^2
\nonumber\\
&&+ 1.203 \times 10^{-1} n^3 - 5.925 \times 10^{-3} n^4 
+ 1.389 \times 10^{-4} n^5.
\eea
The slight difference between the two formulae comes from the difference 
of the values $q'$ (see Fig.~\ref{PDF acceleration log-linear Bodenschatz} 
and Fig.~\ref{PDF acceleration log-linear Gotoh} below).

Within the present stage of the multifractal analysis, the value $q'$ is 
determined by adjusting the center part with the experimental data.
However, since the center part of the PDF takes care only smaller accelerations 
compared with its deviation, we can expect that the dynamical analysis 
such as done by Beck \cite{Beck1,Beck2} may be effective. 
The dynamical study for the center part will be reported elsewhere
in connection with the multifractal characteristics of the system which
determines the tail part representing large deviations.


\section{Analysis of experiments}

\begin{figure}
\begin{center}
\includegraphics[width=14cm]{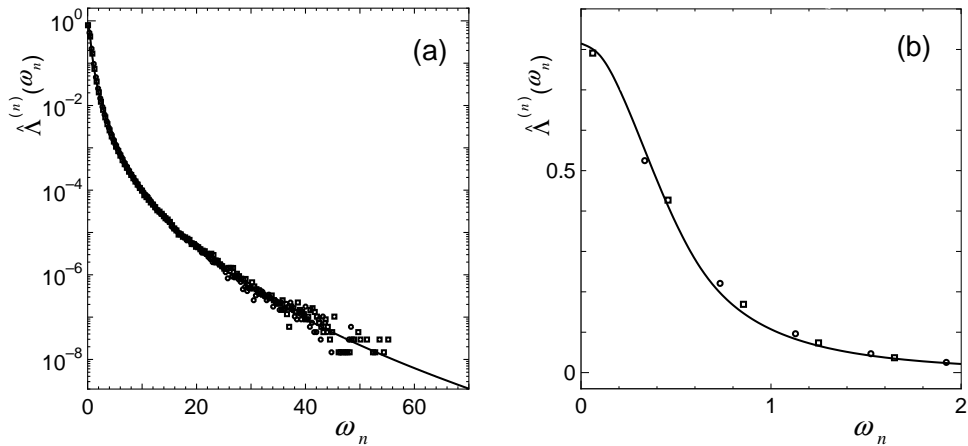}
\end{center}
\caption{PDF of accelerations 
plotted on (a) log
and (b) linear scale.
Comparison between the experimentally measured PDF of 
fluid particle accelerations by 
Bodenschatz et al.\ at $R_\lambda =690$ ($\rRe = 31\ 400$) 
and the present theoretical PDF $\hat{\itLambda}^{(n)}(\omega_n)$.
Open squares are the experimental data points on the left hand side 
of the PDF, whereas open circles are those on the right hand side. 
Solid lines represent the curves given 
by the present theory (\ref{PDF accel large}) and (\ref{PDF accel small})
with $\mu = 0.240$ ($q = 0.391$), $n=17.1$ and $q' = 1.45$.
\label{PDF acceleration log-linear Bodenschatz}}
\end{figure}

We analyze, with the formula (\ref{PDF accel large}) and (\ref{PDF accel small}),
the experimental PDF at $R_\lambda = 690$ measured 
by Bodenschatz et al.\ \cite{EB02comment} 
in Fig.~\ref{PDF acceleration log-linear Bodenschatz} 
on (a) log and (b) linear scale.
We determined the value $n=17.1$ for this experiment
by substituting into its definition (\ref{r-n}), i.e., 
$
n = \log_2 (\ell_0/\ell_n)
$,
the reported value of the integral length scale 0.071~m for $\ell_0$, and of 
the spatial measurement resolution 0.5~$\mu$m for $\ell_n$ \cite{EB01a,EB01b}.
The intermittency exponent $\mu = 0.240$ is extracted by the method of 
least squares with respect to the logarithm of PDFs as the best fit of 
our theoretical formulae with $n=17.1$ to the observed values of 
the PDF \cite{EB01a,EB01b,EB02comment}.
We discarded those points whose PDF values are less than $\sim 10^{-7}$ 
since they scatter largely on log scale.
Substituting the extracted value of $\mu$ into the self-consistent equations, we have 
the values of parameters: $q = 0.391$, $\alpha_0 = 1.138$ and $X = 0.285$.
With these values, other quantities are determined, e.g.,
$
\itDelta \alpha = 1.161
$,
$
\alpha_{+} -\alpha_0 
= \alpha_0 - \alpha_{-} = 0.6815
$,
$
\alpha^\dagger = 1.005
$,
$
\omega_n^\dagger = 0.605
$
and
$
\omega_n^{\rm max} = 2040
$.
The best fit at the center part is given with $q' = 1.45$.\footnote{
It is remarkable that this value of $q'$ is close to $q'=1.5$ proposed by Beck 
\cite{Beck1,Beck2}, although it is claimed that his formula does not 
provide a good explanation of experimental PDF 
for larger accelerations \cite{EB02comment}.
}
The flatness of the PDF of accelerations has the value
$
F_{\mathrm{a}}^{(n)} = 56.9
$
which is within the reported value of the flatness $55\pm4$ \cite{EB02comment}.
This results tells us that the formula (\ref{F-n}) of the flatness can be 
used to derive the value of $\mu$, accurately, when one knows 
the value of $n$, and vice versa 
(see Fig.~\ref{fig: F-n} and also Fig.~\ref{a4Lambda} below).

\begin{figure}
\begin{center}
\includegraphics[width=14cm]{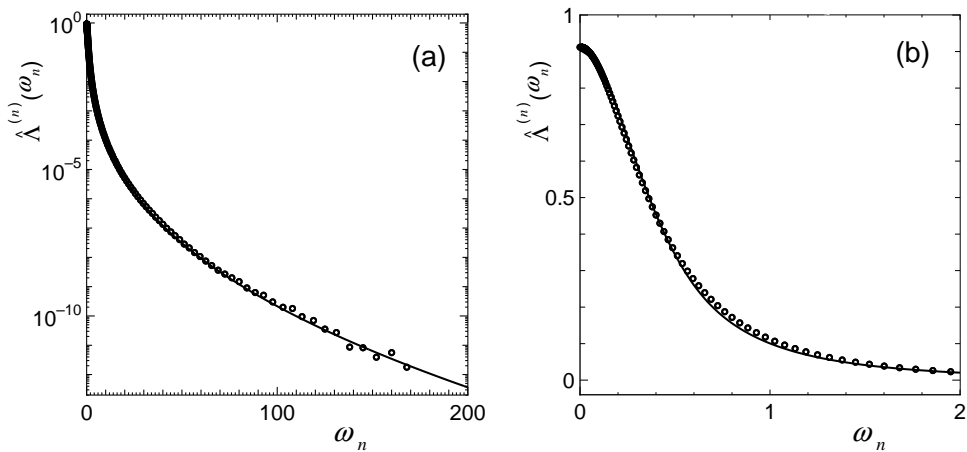}
\end{center}
\caption{PDF of accelerations 
plotted on (a) log
and (b) linear scale.
Comparison between the PDF of fluid particle accelerations 
measured in the DNS by Gotoh et al.\ at $R_\lambda = 380$ 
and the present theoretical PDF $\hat{\itLambda}^{(n)}(\omega_n)$.
Closed circles are the DNS data points both on the left and right hand sides 
of the PDF. Solid lines represent the curves given 
by the present theory (\ref{PDF accel large}) and (\ref{PDF accel small})
with $\mu = 0.240$ ($q = 0.391$), $n=17.5$ and $q' = 1.70$.
\label{PDF acceleration log-linear Gotoh}}
\end{figure}

Analysis of the PDF of accelerations extracted out from the DNS data 
obtained by Gotoh et al.\ \cite{Gotoh02} at $R_\lambda = 380$ is shown 
in Fig.~\ref{PDF acceleration log-linear Gotoh} on (a) log and (b) linear scale.
The value $\mu = 0.240$ for this DNS has been determined by the analysis
of the experimental scaling exponent $\zeta_m$ of longitudinal velocity structure function 
with our theoretical formula (\ref{zeta}) \cite{AA7,AA8}.
Substituting this value of $\mu$ into the self-consistent equations, we have 
the values of parameters: 
$q = 0.391$, $\alpha_0 = 1.138$, $X = 0.285$,
$
\itDelta \alpha = 1.161
$
and
$
\alpha_{+} -\alpha_0 
= \alpha_0 - \alpha_{-} = 0.6815
$
which are the same as those derived in the analysis of the experiment
by Bodenschatz, since both systems have a common value $\mu = 0.240$ for 
the intermittency exponent.
The value $n = 17.5$ is extracted by the method of 
least squares with respect to the logarithm of PDFs as the best fit of 
our theoretical formulae with the derived parameters given above 
to the observed data of the PDF \cite{Gotoh02}.
Note that
$
\alpha^\dagger = 1.005
$,
$
\omega_n^\dagger = 0.622
$
and
$
\omega_n^{\rm max} = 2534
$.
The best fit at the center part is given with $q' = 1.7$. 
The flatness of the PDF is
$
F_{\mathrm{a}}^{(n)} = 70.0
$.
The characteristic distance $r=\ell_n$ for $n=17.5$ reduces to $r/\eta = 7.91$ 
with $\eta \approx 0.258 \times 10^{-2}$ \cite{Gotoh02},
which is about three times longer than the length 
$\Delta r/\eta = 2\pi/(1024 \eta) = 2.38$ of the mesh of Gotoh's DNS.
In deriving the characteristic distance, we used the second formula in\footnote{ 
The empirical formulae (\ref{upper}) and (\ref{lower}) were extracted through 
the multifractal analysis for 
the PDFs of velocity fluctuations measured at two points separated 
by $r$ \cite{AA7}.
The corresponding numbers $n$ and $\bar{n}$ are given by
($r/\eta$, $n$, $\bar{n}$) = (2.38, 21.5, 14.5), (4.76, 20.0, 13.0), 
(9.52, 16.8, 9.81), (19.0, 14.0, 7.01), (38.1, 11.8, 4.81), 
(76.2, 10.1, 3.11), (152, 9.30, 2.31), (305, 8.10, 1.11),
(609, 7.00, 0.01), (1220, 6.00, -0.99).
It seems that there are two scaling regions constituting the inertial range. 
The formula (\ref{lower}) indicates that 
in the lower scaling region ($r < \ell_{\rm c}$)
eddies break up, effectively, into 1.33 pieces in contrast to 
the upper scaling region ($\ell_{\rm c} \leq r$)
where eddies break up into two pieces as seen in (\ref{upper}) \cite{AA7}. 
The existence of the two scaling regions can be an artifact of DNS 
due to a shortage of calculation time, since
energy cascading process is not so effective for eddies whose sizes are
smaller than $\ell_{\rm c}$ or the Taylor microscale.
Note that Gotoh et al.\ claimed that the inertial range is restricted
only to the region around $\bar{n}=$ 3.11, 2.31 and 1.11 \cite{Gotoh02}. 
}
\bea
n \aeq - 1.050 \log_2 (r/\eta) +16.74 \quad (\ell_{\rm c} \leq r)
\label{upper}
\\
n \aeq - 2.540 \log_2 (r/\eta) +25.08 \quad (r < \ell_{\rm c}).
\label{lower}
\eea
Here, $\ell_{\rm c}/\eta = 48.26$ is the crossover length of two scaling regions, 
and is close to the Taylor microscale $\lambda/\eta = 38.33$.
With the help of the formula (\ref{n-nbar}), we have $\bar{n}=11.8$.

\begin{figure}
\begin{center}
\includegraphics[width=12cm]{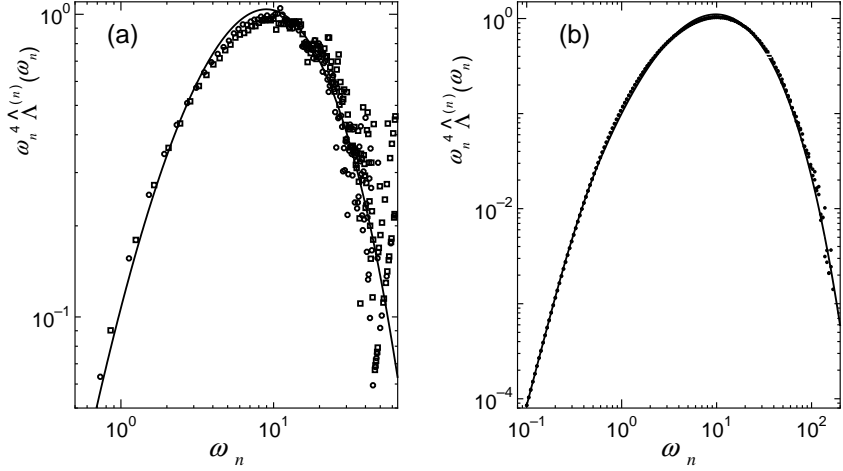}
\end{center}
\caption{Comparison of the theoretical curves (solid lines) for 
the integrand $\omega_n^4 \hat{\itLambda}^{(n)}(\omega_n)$ of the flatness
$F_{\mathrm{a}}^{(n)}$ with the corresponding
(a) experimental quantity by Bodenschatz et al.\ 
(open squares for the data on the left hand side, 
and open circles on the right hand side), 
and (b) DNS quantity by Gotoh et al.\ (closed circles both for the data 
on the left and right hand) in loglog scale.
The parameters in the theoretical PDFs for (a) and (b) are 
the same as those given in Fig.~\ref{PDF acceleration log-linear Bodenschatz} and
Fig.~\ref{PDF acceleration log-linear Gotoh}, respectively.
\label{a4Lambda}}
\end{figure}

In Fig.~\ref{a4Lambda}, we put the lines representing the integrand 
$\omega_n^4 \hat{\itLambda}^{(n)}(\omega_n)$ of the flatness $F_{\mathrm{a}}^{(n)}$ 
and the corresponding experimental data (a) by Bodenschatz et al.\ 
\cite{EB02comment}, and (b) by Gotoh et al.\ \cite{Gotoh02}. 
The agreements are remarkable.

\section{Comment on the energy-input scale}

There is no room to incorporate, {\em automatically}, 
into the present multifractal analysis
the energy input scale $\ell_{\rm in}$ and the "system size" $\ell_0$.
The former is necessary to determine 
the number of steps $\bar{n}$ in the energy cascade model.
Once $\ell_{\rm in}$ is determined by investigating the structure of 
experimental apparatuses, the relation between $\bar{n}$ and 
the multifractal step $n$ is given by (\ref{n-nbar}).
Since main part of the multifractal analysis rests on the scale invariance,
the size of the system under consideration is assumed to be infinite, and 
therefore, the value of the reference length $\ell_0$ introduced in (\ref{r-n}) 
is determined only through the analysis of experimental data.

Actually, for example, the empirical equation \cite{AA5}
\be
n = -1.019 \log_2 r/\eta + 0.901 \log_2 \rRe
\label{n-roeta experimental}
\ee
extracted from the experimental PDFs for velocity fluctuations\footnote{
The velocity fluctuations are measured at two separated points whose distance is $r$.
The corresponding numbers $n$ and $\bar{n}$ are given by \cite{AA9}
($r/\eta$, $n$, $\bar{n}$) = (11.6, 14, 10.7), (23.1, 13, 9.7), (46.2, 11, 8.7),
(92.5, 10, 7.7), (208, 9.0, 6.6), (399, 8.0, 5.6), (830, 7.5, 4.6), (1440, 7.0, 3.8).
The relation (\ref{n-roeta experimental}) is extracted with these values \cite{AA5}.
}
by Lewis and Swinney \cite{Lewis-Swinney99} 
gives 
$
\ell_0 \approx 877 \mbox{ cm}
$.
In their analysis, the Reynolds number $\rRe = 540\ 000$
is estimated with 
$
\ell_{\rm in} = 2 \pi \times 19.00 \mbox{ cm} \approx 119.32 \mbox{ cm}
$
and the Kolmogorov scale 
$
\eta \approx 0.006 \mbox{ cm}
$. 
Therefore, the reference length is large compared with the energy-input scale.

For Gotoh's DNS, the empirical equation (\ref{upper}), extracted by 
the analysis of PDFs of velocity fluctuations, gives 
$
\ell_0 /\eta \approx 63\ 000
$
which is larger than the energy-input scale 
$
\ell_{\rm in} /\eta = \pi/(k\eta) \approx 497
$
with the wavenumber $k= 6^{1/2}$ of forcing 
where we took 209 grid point spacings for the measure
of the energy-input scale \cite{Gotoh-Kraichnan03}.
We see again that the reference length is longer than the energy-input scale.
The Reynolds number of the DNS is now estimated by (\ref{Re}) as ${\rm Re} = 3\ 937$.
A unified study of Gotoh's DNS~\cite{Gotoh02}
by means of the multifractal analysis with the Tsallis-type PDF for
the center part will be given elsewhere,
in which it is shown that the PDFs for the velocity fluctuations, 
for the velocity derivatives and for the fluid particle accelerations 
in addition to the scale exponents of velocity structure function
provide us with consistent results.

In the analysis of Bodenschatz's experiment, 
the identification of $\ell_0$ with the integral length scale 7.1~cm gives us 
reasonable value of $n=17.1$ in the sense that with this value 
the formula (\ref{F-n}) of the flatness provides us with $\mu =0.240$
for the intermittency exponent.
This value turns out to be the same as the one observed by Gotoh et al.\ in their DNS.
Since Bodenschatz assigned $\rRe = 31\ 400$ with the formula (\ref{Re}), 
$\eta = 30.3$~$\mu$m gives $\ell_{\rm in} = 0.071$~m, as it should be.
Therefore, in this case, we see that the energy-input scale is equal to
the integral length scale, i.e., $\ell_{\rm in} = \ell_0$, and that
$\bar{n} = n = 17.1$.

\section{Discussions}

\begin{figure}
\begin{center}
\includegraphics[width=14cm]{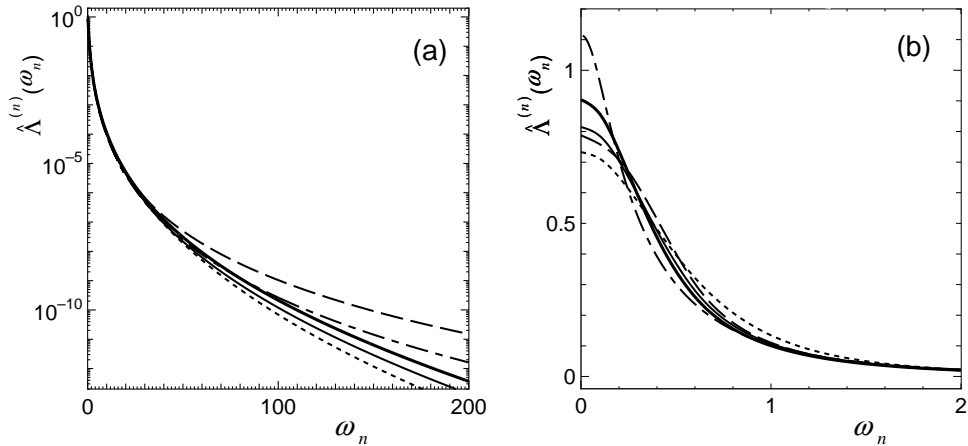}
\end{center}
\caption{Comparison of PDFs of accelerations on (a) log and (b) linear scale.
The thin and thick lines are, respectively, the present theoretical PDFs 
in Fig.~\ref{PDF acceleration log-linear Bodenschatz} 
and Fig.~\ref{PDF acceleration log-linear Gotoh}.
The dashed line represents the PDF within the log-normal model,
and the dotted line does the empirical PDF given by Bodenschatz et al..
The dotted-dashed line is the PDF given by Beck \cite{Beck3}.
\label{PDF comparison}}
\end{figure}

A comparison of the PDFs of accelerations are put in Fig.~\ref{PDF comparison} 
on (a) log and (b) linear scale.
The thin and the thick lines are those theoretical PDFs 
in Fig.~\ref{PDF acceleration log-linear Bodenschatz}
and in Fig.~\ref{PDF acceleration log-linear Gotoh}, respectively.
The analysis of the data by Bodenschatz for the accelerations
in terms of the log-normal model was performed just the same way
as the multifractal analysis to get $\mu = 0.260$ and $n = 15.5$.
The resulting PDF is given by dashed line which deviates slightly upward
for larger values of $\omega_n$ compared with the thin line. 
This deviation can be understood from 
the scaling exponents $\zeta_m$ of the velocity structure function 
within the log-normal model. It becomes negative for larger values of $m$.
The dotted line is the empirical PDF given by Bodenschatz et al.\ 
\cite{EB02comment} which deviate downward for larger values of $\omega_n$.
The dotted-dashed line is the PDF given by Beck \cite{Beck3} for 
the studies both of the PDFs by Bodenschatz et al.\ and of
Gotoh et al..
The lines other than the present theoretical PDFs in Fig.~\ref{PDF comparison}
give rather poor explanations at the center part.\footnote{
Beck introduced a somewhat artificial cutoff in his formula in order to fix
its defect at the center part of the PDF \cite{Beck4}.
}

From the above analyses of two experiments, we reveal that there are
two mechanisms contributing to the PDF of the accelerations, i.e., one is
for the tail part, and the other for the center part.
The structure of the PDF $\hat{\itLambda}^{(n)}(\omega_n)$ for the tail part,
$\omega_n^\dagger \leq \vert \omega_n \vert$, is given by (\ref{PDF accel large})
that represents the intermittent large deviations 
which is a manifestation of the multifractal distribution of singularities 
in physical space due to the scale invariance of 
the Navier-Stokes equation for large Reynolds number. 
The experiment conducted by Bodenschatz et al.\ \cite{EB01a,EB01b} visualized 
the singularities in physical space by tracing the "fluid particle" in turbulence.
The specific form (\ref{PDF accel large}) comes from 
the distribution function for the singularity exponent $\alpha$ that is represented
by the Tsallis-type distribution function (\ref{Tsallis prob density}) 
with the parameter $q$ which is
determined by the observed value of the intermittency exponent.
The flatness of the PDF mainly provides us with the information of this tail part.
The structure of the PDF for the center part, 
$\vert \omega_n \vert \leq \omega_n^\dagger$, is given by (\ref{PDF accel small})
that represents small deviations violating the scale invariance
due to thermal fluctuations and/or observation error.
The PDF for this part is assumed to be given by the Tsallis-type distribution function 
for acceleration itself with the parameter $q'$.
In this paper, the value of $q'$ are determined with the help of 
the experimentally observed PDF at the center part, giving $q'=1.45$ and $q'=1.7$
which is close to 1.5 proposed by Beck \cite{Beck1,Beck2}.
The value of $q'$ should be determined by investigating the dynamics of
thermal fluctuations and/or observation error on the multifractal support,
which may have non-additive character.
This is one of the attractive future problems.
Note that we already saw a dependence of $\ln (1/(q'-1))$ on $\ln(r/\eta)$
through the study of the PDFs of velocity fluctuations,
a detailed of which will be reported elsewhere.
The tail part and the center part are separated at $\omega_n^\dagger \approx 0.6$.
This gives $\alpha^\dagger \approx 1$ that satisfies the condition $\alpha < 1.5$ 
in which the singularity appears in fluid particle accelerations.

We saw that the PDFs derived within the multifractal analysis seem to be sensitive to
the characteristic lengths such as the distance of two measuring points,
the space resolution in measurement and the mesh size of DNS.
We also knew that the multifractal distribution of singularities in physical space, 
on which the present analysis rests, is robust enough to allow us
to apply the multifractal analysis to the ranges outside of the inertial range.
How to put the information of energy-input scale
into the multifractal analysis is one of the important future problems 
\cite{GotohKraichnan03}. 
It may be resolved when one succeeds to reveal the dynamical foundation
underlying the basis of the multifractal analysis, starting an investigation by 
the stochastic Navier-Stokes equation with the energy input term.
If one could put the stochastic Navier-Stokes equation 
under the influence of white Gaussian noise, describing thermal fluctuation related to 
kinematic viscosity, into a linealized stochastic equation with 
a renormalized turbulent viscosity \cite{Heisenberg}, 
the relevant stochastic process should be the one related to 
the PDF of accelerations derived in this paper, which may be named
{\it R\'enyi} or {\it Tsallis process}.
The success of the multifractal analysis in the
studies of experimental PDFs may indicate that the multifractal
distribution of singularities in physical space are robust against the addition of
the energy-input term.

The application of the present multifractal analysis to the micrometeorological
study of atmospheric turbulence is one of the attractive future problems.
The tail part is specified by the intermittency exponent, and the center part may
give us the information, for example, about the thermodynamical structure of 
the canopy in Amazon forest \cite{Ramos03}.
The application to the vortex tangle \cite{Feynman} in the superfluids $^4$He and 
$^3$He is another examples of the exciting future problems.
In low temperature, the tail part of PDF comes from the singularity 
in the superfluid component within the two fluid model, and is determined by 
the intermittency exponent. 
Whereas, the value of $q'$ for the center part of PDF may 
be determined by the dynamical structure of quantized vortices.
The multifractal analysis may also open new aspect for the systems with large deviation
found in a large variety of areas (see, for example, the web site in \cite{Tsallis99}),
and its applications to these areas will be reported elsewhere.

\section*{Acknowledgements}

The authors would like to thank Prof.~R.H.~Kraichnan for his critical and 
enlightening comments, and Prof.~C.~Tsallis for his fruitful comments 
with encouragement.
The authors are grateful to Prof.~E.~Bodenschatz and Prof.~T.~Gotoh
for the kindness to show their data prior to publication.

\end{document}